\documentclass{article}

\newcommand{\lw}[1]{\smash{\lower2.ex\hbox{#1}}}
\newcommand{\lwo}[1]{\smash{\lower1.ex\hbox{#1}}}
\newfont{\bg}{cmr10 scaled\magstep4}
\newcommand{\bigzero}{\mbox{\boldmath $0$}}
\newcommand{\bigzerol}{\smash{\hbox{\bg 0}}}
\newcommand{\bigzerou}{\smash{\lower1.7ex\hbox{\bg 0}}}
\usepackage[dvips]{graphicx}
\usepackage{fancybox}

\usepackage[lflt]{floatflt}
\begin{document}
\title{\Large\bf General-Purpose Parallel Simulator for Quantum 
        Computing}
\author{
{\large \hspace*{-1ex} $\mbox{\bf Jumpei Niwa}^{\ast}$ \hspace*{-
1ex}}\\
{\tt \hspace*{-2ex} niwa@is.s.u-tokyo.ac.jp \hspace*{-2ex}} 
\and
{\large \hspace*{-1ex} $\mbox{\bf Keiji Matsumoto}^{\dagger}$ \hspace*{-1ex}}\\
{\tt \hspace*{-2ex} keiji@qci.jst.go.jp \hspace*{-2ex}}
\and
{\large \hspace*{-1ex} $\mbox{\bf Hiroshi Imai}^{\ast \dagger}$ \hspace*{-1ex}}
\\
{\tt \hspace*{-2ex} imai@is.s.u-tokyo.ac.jp \hspace*{-2ex}}
}

\date{\today}
\maketitle
\footnotetext[1]{
\noindent
Department of Computer Science, Graduate School of Information Science
and Technology, The University of Tokyo, 7-3-1 Hongo, Bunkyo-ku, Tokyo
113-0033, Japan.}
\footnotetext[2]{
\noindent
Quantum Computation and Information Project, ERATO,
Japan Science and Technology Corporation,
5-28-3 Hongo, Bunkyo-ku, Tokyo 113-0033, Japan.}

\begin{abstract}
With current technologies, it seems to be very difficult to implement
quantum computers with many qubits. It is therefore of importance to
simulate quantum algorithms and circuits on the existing computers.
However, for a large-size problem, the simulation often requires more
computational power than is available from sequential
processing. Therefore, the simulation methods using parallel
processing are required.

We have developed a general-purpose simulator for quantum computing on
the parallel computer (Sun, Enterprise4500). It can deal with up-to
\emph{30 qubits}. We have performed Shor's
factorization and Grover's database search by using the simulator, and
we analyzed robustness of the corresponding quantum circuits in the
presence of decoherence and operational errors. The corresponding results, statistics and
analyses are presented.
\vskip 1em
\noindent
{\bf key words} : quantum computer simulator, Shor's factorization,
  Grover's database search, parallel processing, decoherence and
  operational errors
\end{abstract}

\section{Introduction}
With the current technologies, it seems to be very difficult to implement
quantum computers with many qubits. It is therefore of importance to
simulate quantum algorithms and circuits on the existing computers.
The purpose of the simulation is
\begin{itemize}
\item to investigate quantum algorithms behavior.
\item to analyze performance and robustness of quantum circuits in
  the presence of decoherence and operational errors.
\end{itemize}
However, simulations often require more computational power than
is usually available on sequential computers. Therefore, we have developed
the simulation method for parallel computers. That is, we have
developed a general-purpose simulator for quantum algorithms and
circuits on the parallel computer, Symmetric Multi-Processor.

\begin{figure}[hbtp]
  \vspace*{-0.5cm}
  \begin{center}
    \leavevmode \includegraphics*[width=0.33\textwidth]{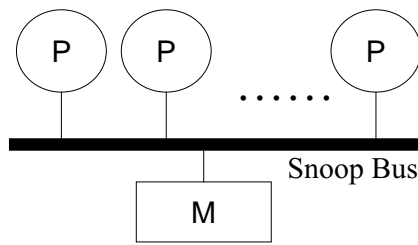}\\
    P: Processor, M: Memory
    \vspace*{-0.2cm}
    \caption{SMP (Symmetric Multi-Processors).}
  \end{center}
  \vspace*{-0.5cm}
\end{figure}

\section{Basic Design}
\subsection{Registers}

The simulation is for quantum circuit model of computation. 
A collection of $n$ qubits is called a \emph{register} of size $n$. 
The general qubit state of the $n$-qubit register is
\[ | \phi \rangle = \sum_{i=0}^{2^n-1} \alpha_i | i \rangle \mbox
{ where } \alpha_i \in \mathcal{C} \mbox{ , }\sum_{i=0}^{2^n-1}
|\alpha_i|^2 = 1\mbox{ .}\] That is, the state of an $n$-qubit register is
represented by a unit-length complex vector 
on $\mathcal{H}_{2^n}$. In a classical computer, to store a complex
number $\alpha = x + i y$, one require to store a pair of real numbers
$(x, y)$. Each real number will be represented by a {\it double
  precision word}. The double precision word is 16 bytes (64bits) on
most of the systems. $2^{n+4}$ bytes memory is therefore required to
deal with the state of an $n$-qubit register in a classical computer.

\subsection{Evolution}
The time evolution of an $n$-qubit register is determined by a unitary
operator on $\mathcal{H}_{2^n}$. The size of the matrix is $2^n \times
2^n$.  In general, it requires $2^n \times 2^n$ space and
$2^n(2^{n+1}-1)$ arithmetic operations to perform classically such an
evolution step.

However, we mostly use operators that have simple structures when we
design quantum circuits. That is, an evolution step is performed by
applying a unitary operator ($2\times2$) to a single qubit (a single
qubit gate) or by applying the controlled unitary operator such as a
C-NOT gate. It requires only $2\times2$ space and $3\cdot2^{n}$
arithmetic operations to simulate such an evolution step.

\subsubsection{A Single Qubit Gate}
\label{sq}

\begin{floatingfigure}{4.5cm}
  \vspace*{-0.1cm}
  \begin{center}
    \includegraphics*[width=3.8cm]{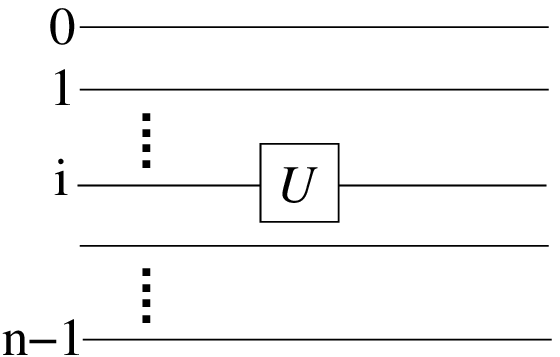}\\
    \vspace*{-0.3cm}
    \caption{Single qubit gate.}
  \end{center}
\end{floatingfigure}
Suppose that the MSB (most significant bit) is $0$-th qubit.
When a unitary matrix
$\mbox{\boldmath $U$} = \left(
  \begin{array}{cc}
    u_{11} & u_{12} \\
    u_{21} & u_{22}
  \end{array}
\right)$ is applied to the $i$-th qubit,  the overall unitary
operation applied to the $n$-qubit register state has the form $X =
(\bigotimes_{k=1}^{i-1}I) \allowbreak \bigotimes U \bigotimes
(\bigotimes^{n}_{k=i+1} I)$. $2^n \times 2^n$ matrix $X$ is the sparse
regular matrix shown in Figure \ref{fig:tum}.

\begin{figure}[htbp]
  \begin{center}
    \leavevmode
    \[X = \underbrace{\left(
      \begin{array}{cccccc}
        S_{0} &   & & && \bigzerou \\
          & S_{1} & & & &\\
          & & \cdots & & & \\
          & & & \cdots & & \\
          & & & & S_{2^i-2} & \\
         \bigzerol & & & & & S_{2^i-1}\\
      \end{array}
      \right)}_{2^n} \mbox{ where } S_k =
    \underbrace{
    \left(\begin{array}{cccccc}
          u_{11} & & \bigzero & u_{12} & & \bigzero \\
          & \cdots & & & \cdots & \\
          \bigzero & & u_{11} & \bigzero & & u_{12} \\
          u_{21} & & \bigzero & u_{22} & & \bigzero \\
          & \cdots & & & \cdots & \\
          \bigzero & & u_{21} & \bigzero & & u_{22} \\
      \end{array}
      \right)}_{2^{n-i}} \mbox{ }(0\le k < 2^{i})
\]
    \caption{Total unitary matrix.}
    \label{fig:tum}
  \end{center}
\end{figure}

We therefore do not have to generate $X$ explicitly. We have only to
store the $2 \times 2$ matrix $U$. Since there are only 2 non-zero
elements for each row in $X$, the evolution step (i.e., multiply of a
matrix and a vector) is simulated in $3 \cdot 2^n$ arithmetical
operations. 

\paragraph{Parallelization}
~\\
Of course, the evolution step ($X |\phi \rangle$) can
be executed in parallel. Let $2^P$ be the number of processors
available in the simulation system. The evolution step is decomposed
into a sequence of submatrix-subvector multiplication $M_k$ ($0 \le k
< 2^i$). $M_k$ is defined as $S_k \phi_k$, that is, the multiplication
of a submatrix $S_k$ ($2^{n-i} \times 2^{n-i}$) and a subvector
$\phi_k$ whose length is $2^{n-i}$ (shown in Figure \ref{fig:pe}).
Note that there are no data-dependencies between $M_k$ and $M_l$ ($k
\ne l$). Therefore, $M_k$ and $M_l$ are executed in parallel.  We
assign $\underbrace{M_{p2^{i-P}},M_{p2^{i-P}+1},\ldots,
  M_{(p+1)2^{i-P}-1}}_{2^{i-P}}$  to a processor $p$ ($0\le p <
2^P$). That is, the processor $p$ computes $2^{i-P}$
submatrix-subvector multiplications, and the rests of multiplications
are performed in other processors in parallel. After each processor
has finished its assigned computations, it executes a synchronization
primitive, such as the barrier, to make its modifications to the
vector ($\phi$), that is, the state of the register visible to other
processors.

\begin{figure}[htbp]
  \begin{center}
    \leavevmode
    \[X|\phi \rangle = \left(
      \begin{array}{cccccccc}
        S_{0} &   & & & & && \bigzerou \\
          & S_{1} & & & & & &\\ 
          & & \cdots & & & & & \\ \hline
          & & & \cdots & & & & \\
          & & & & \cdots & & & \\ \hline
          & & & & & \cdots & & \\ 
          & & & & & & S_{2^i-2} & \\
         \bigzerol & & & & & & & S_{2^i-1}\\
      \end{array}
      \right)
      \left(
        \begin{array}{c}
        \phi_{0}\\
        \phi_{1}\\
        \cdots\\ \hline
        \cdots\\
        \cdots\\ \hline
        \cdots\\ 
        \phi_{2^i-2} \\
        \phi_{2^i-1}\\
      \end{array}
    \right)
    \hspace*{-0.5cm}
      \begin{array}{c}
        \left.
        \begin{array}{c}
          \\
          \\
        \end{array}\hspace*{-0.3cm}\right\}(processor~0)\\ 
      \cdots\\
      \cdots\\
      \cdots\\
        \left.
          \begin{array}{c}
          \\
          \\
          \end{array} \right\} (processor~2^P)\\
        \end{array}
      \begin{array}{c}
        \\
        \hspace*{-0.5cm}\mbox{where } 
      \phi_k =
      \left(\begin{array}{c}
          \alpha_{k2^{n-i}}\\
          \alpha_{k2^{n-i}+1}\\
          \cdots\\
          \alpha_{(k+1)2^{n-i}-2}\\
          \alpha_{(k+1)2^{n-i}-1}\\
      \end{array}
      \right) \\ \\
      \hspace*{-2cm}(0 \le k < 2^i)
      \end{array}
\]
    \caption{Computation decomposition in the general case.}
    \label{fig:pe}
  \end{center}
\end{figure}

\begin{figure}[htbp]
  \begin{center}
    \leavevmode \includegraphics*[width=0.7\textwidth]{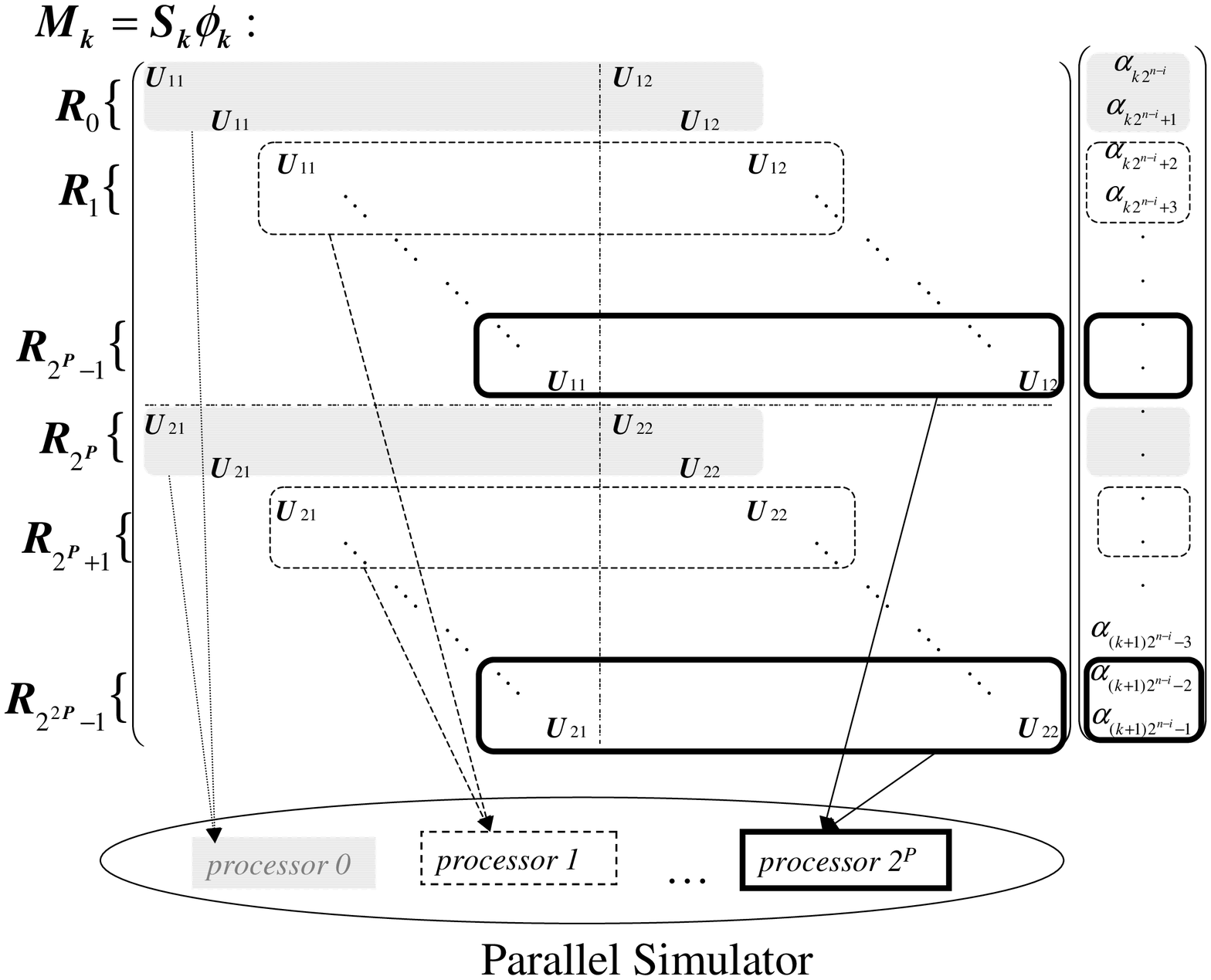}
    \caption{Computation decomposition in the large subblock case.}
    \label{fig:sub}
  \end{center}
\end{figure}

When the number of submatrices is smaller than the number of
processors (i.e., $2^i < 2^P$), it is inefficient to
assign the computation $M_k(= S_k\phi_k$, $0\le k< 2^i$)) to one processor
as described above. It can cause a load imbalance in the simulation
system. In this case, we should decompose the computation $M_k$
itself to improve parallel efficiency. Each submatrix $S_k$ is divided
into $2^{P+1}$ chunks of rows. Each chunk of rows $R_j$ $(0 \le j <
2^{P+1})$ contains the contiguous $2^{n-i-(P+1)}$ rows of $S_k$. The
multiplications using the chunk of rows $R_j$ and $R_{2^P+j}$ are
assigned to a processor $j$ as described in the Figure
\ref{fig:sub}. This decomposition is applied to all the $M_k$
computations ($0\le k< 2^i$). 

Note that the computation using $j$-th row of the submatrix must be
always paired with that using  $(j+2^{n-i-1})$-th row when we use an
``in-place'' algorithm (i.e., The results of $X|\phi\rangle$ are
stored in $|\phi\rangle$). That is, multiplications using the chunk of
rows $R_j$ and $R_{2^P+j}$ are assigned to the same processor $j$.
This is because there are dependencies across processors. Consider the
following example.

\[
\hspace*{-0.7cm}
\left[\begin{array}{c}
    xu_{11}+yu_{12}\\
    \ldots\\
    \ldots\\
    xu_{21}+yu_{22}\\
    \ldots\\
    \ldots
  \end{array}\right] 
= \left[\begin{array}{cccccc}
    u_{11} & & \bigzero & u_{12} & & \bigzero \\
    & \ldots &  & & \ldots & \\
    \bigzero &  & u_{11} & \bigzero & & u_{12} \\
    u_{21} & & \bigzero & u_{22} & & \bigzero \\
    & \ldots &  & & \ldots & \\
    \bigzero &  & u_{21} & \bigzero & & u_{12} \\
  \end{array}\right] 
\left[\begin{array}{c}
    x\\
    \ldots\\
    \ldots\\
    y\\
    \ldots\\
    \ldots
  \end{array}\right] 
\]
If the 1-st element is computed and the result ($xu_{11}+yu_{12}$) is
stored before the 4-th element is computed, the result of 4-th element
computation becomes not $xu_{21}+yu_{22}$ but
$(xu_{11}+yu_{12})u_{21}+yu_{22}$. This is wrong. To avoid this
situation, all the processors have only to execute barrier operations
before storing the computed results. However, a barrier operation per
store operation can cause heavy overheads.

Therefore, the 1-st element computation and 4-th element computation
should be assigned to the same processor. Then, the data-dependencies
are not cross-processor but in-processor. First, the processor
computes $xu_{11}+yu_{12}$ and stores the result in a temporary
variable $t_1$ on the local storage-area (i.e., stack). Second, the processor itself
computes the result $xu_{21}+yu_{22}$ and stores it in the 4-th
element. Third, the processor stores the contents of the temporary
variable $t_1$ in the 1-st element. In this way, we can avoid the
above wrong situation without performing synchronization primitives.
If there are no overheads for parallel execution, the time complexity
is thus reduced to $O(2^{n-P})$ where $2^P$ is the number of
processors available in the system.

\subsubsection{A Controlled Qubit Gate}
\begin{floatingfigure}{4.3cm}
  \vspace*{-0.3cm}
  \begin{center}
    \includegraphics*[width=3.8cm]{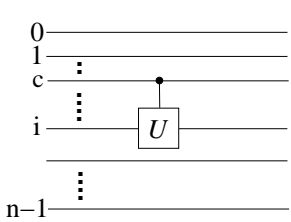}\\
    \vspace*{-0.3cm}
    \caption{Controlled qubit gate.}
  \end{center}
\end{floatingfigure}

Suppose that a unitary matrix $\mbox{\boldmath $U$} = \left(
  \begin{array}{cc}
    u_{11} & u_{12} \\
    u_{21} & u_{22}
  \end{array} \right)$ is applied to the $i$-th qubit if and only if
the $c$-th bit (controlled bit) is $1$.  Let $CTX$ be the overall
unitary matrix ($2^n \times 2^n$).  First, we consider the matrix $X$
mentioned in Sec.\ref{sq} as if there were no controlled bits. 
Then,
for each $j$ $(0 \le j < 2^n -1)$, the $j$-th row of $CTX$ ($CTX[j]$)
is defined as follows.

\[ CTX[j] = \left\{ 
  \begin{array}{rl}
    X[j] & \mbox{the $c$-th bit in $j$ is $1$}\\
    I[j] & \mbox{the $c$-th bit in $j$ is $0$}
  \end{array}\right.
\]

where $I$ is the unit matrix. In this case, we also do not have to
generate $CTX$ or $X$ explicitly. We have only to store the $2 \times
2$ matrix $U$. In many controlled bit cases, it is easy to extend this
method. The evolution step is executed in parallel as described in Sec
\ref{sq}. Therefore, the simulation time is $O(2^{n-P})$ when there
are no overheads for parallel execution ($2^P$ is the number of
processors available in the simulation system.)

The simulator provides a \emph{f-controlled U gate}. It is similar to
the controlled $U$ gate. The $U$ gate is applied to the target bit iff
$f(c) = 1$ (the $c$-th bit is the controlled bit). It is used in the
Grover's Search Algorithm \cite{grover96fast}.

\subsubsection{Measurement Gates}
The measurement step 
for an $n$-qubit register state is simulated in $O(2^n)$ time as
follows.
Let $ | \phi \rangle = \sum_{j=0}^{2^n-1} \alpha_j | j \rangle $ be
an $n$-qubit register state.
\begin{enumerate}
\item Generate a random number $r$ ($0 \le r < 1$)
\item Determine an integer $i$ ($0 \le i \le 2^n -1$), s.t.
  \[ \sum_{j=0}^{i-1} |\alpha_j|^2 \le r < \sum_{j=0}^{i}
  |\alpha_j|^2\]
\end{enumerate}
We consider that the measurement is done with respect to the
standard basis $|i \rangle $.

\newpage
\subsection{Basic Circuits}
\subsubsection{Hadamard Transform}
\label{subsec:HT}
\begin{floatingfigure}{4cm}
  \begin{center}
    \vspace*{-0.1cm}
    \includegraphics*[width=2cm]{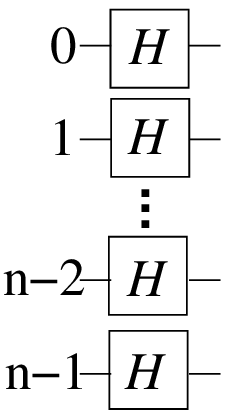}\\
    \vspace*{-0.3cm}
    \caption{Hadamard circuit.}
    \label{fig:hadamard}
  \end{center}
\end{floatingfigure}

The Hadamard transform $H_n$ is defined as follows,
\begin{displaymath}
H_n|x\rangle = \frac{1}{\sqrt {2^n}} \sum_{y \in {0, 1}^n}
(-1)^{x\cdot y} |y\rangle,  
\end{displaymath}
$\mbox{for } x \in \{0, 1\}^n.$
$H_n$ is implemented by the circuit in Figure \ref{fig:hadamard},
where $\mbox{\boldmath $H$}$ denotes ${\displaystyle \frac{1}{\sqrt
    2}\left(
    \begin{array}{cc}
      1 & 1 \\
      1 & -1
    \end{array}
  \right) }$ .
Note that it requires $O(n2^{n-P})$ time when there are no overheads
for parallel execution ($2^P$ is the number of processors available in
the simulation system.).

\vspace*{0.2cm}
\subsubsection{Quantum Fourier Transform}

The quantum Fourier transform (QFT) is a unitary operation that
essentially performs the DFT on quantum register states.  The QFT
maps a quantum state $| \phi \rangle = \sum_{x=0}^{2^n-1} \alpha_x |
x \rangle $ to the state $\sum_{x=0}^{2^n-1} \beta_x | x \rangle $,
where
\[ \beta_x = \frac{1}{\sqrt {2^n}}\sum_{y=0}^{2^n-1}
\omega^{xy}\alpha_y \mbox{,\quad } \omega = e^{2\pi i/2^n}\]

\begin{figure}[htbp]
  \begin{center}
    \includegraphics*[width=0.7\textwidth]{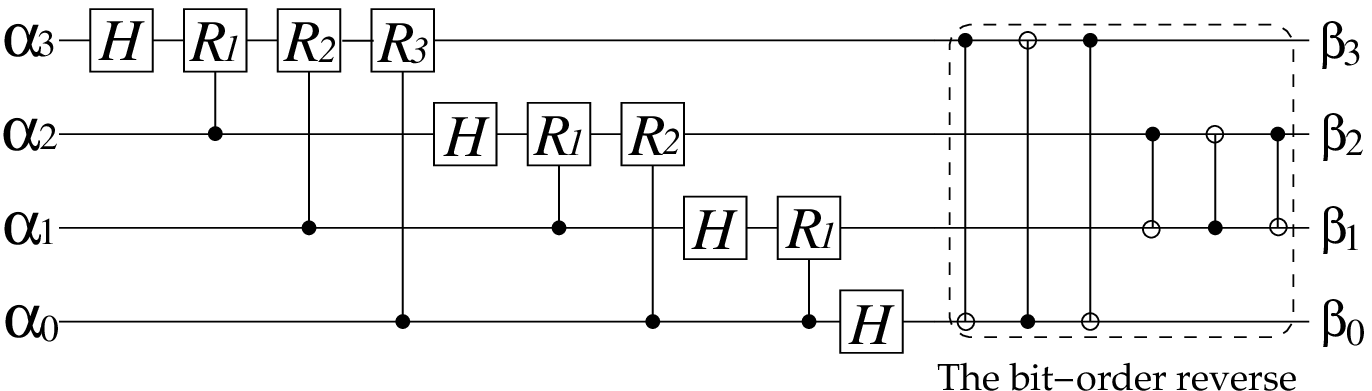}
    \caption{The $\mathrm{QFT}_{2^n}$ circuit ($n=4$).}
    \label{fig:qft}
  \end{center}
\end{figure}

The circuit implementing the QFT is described in the Figure
\ref{fig:qft}. $\mbox{\boldmath $H$}$ is the Hadamard gate, and
$\mbox{\boldmath $R_d$}$ is the phase shift gate denoted as $\left(
  \begin{array}{cc}
    1 & 0 \\
    0 & e^{i\pi/2^d}
  \end{array}
\right)$.

For general $n$, this circuit has $O(n^2)$ size\footnote{There is a
  quantum circuit that computes QFT (modulo $2^n$) that has the size
  $O(n(\log n)^2 \log \log n)$ \cite{cleve00fast}}.  Therefore, the
evolution step is simulated in $O(n^2 2^{n-P})$ time when there are no 
overheads for parallel execution (There are $2^P$ processors available 
in the system). Of course, we can reduce the circuit size to $O(n \log
(n / \mbox{$\mit\epsilon$}))$ \cite{barenco96approximate, cleve00fast}
if we settle the implementation of fixed accuracy ($\mit\epsilon$),
because the controlled phase shift gates acting on distantly separated
qubits contribute only exponentially small phases. In this case, the
evolution step is simulated in $O(n
\log(n/\mbox{$\mit\epsilon$})2^{n-P})$ when there are no overheads for 
parallel execution.

If we regard the QFT transform as a {\em black box operator} (that is,
if we suppose that this QFT circuit has no error), we do not have to
use this quantum circuit in the simulator to perform QFT
transformation.  We can use fast Fourier transform (FFT) in the
simulator instead of the QFT circuit. The FFT algorithm requires only
$O(n 2^{n - P})$ steps when there are no overheads for parallel
execution. Of course, the FFT gives the exact solution.  We use the
8-radix in-place FFT algorithm.

\subsubsection{Arithmetical circuits} 
\label{sec:ari}
The arithmetical circuits are important for quantum computing
\cite{vedral95quantum}. In the Shor's factoring
algorithm\cite{shor94algorithms}, the arithmetical circuits to compute
modular exponentiation are used. Therefore, according to Ref
\cite{miquel96factoring}, we have implemented the modular
exponentiation circuit by using constant adders, constant modular
adders and constant multipliers. $x^a(\bmod \ N)$ can be computed
using the decomposition,
\[ x^a (\mathrm{mod} \mbox{ }N) = \prod_{i=0}^{l-1} \left( (x^{2^i})^{a_i}
  (\mathrm{mod}\mbox{ }N)\right),\ a = \sum_{i=0}^{l-1} a_i 2^i
(=a_{l-1}a_{l-2}\ldots a_{0}\mbox{ (binary representation)})\]
Thus, modular exponentiation is just a chain of products where each
factor is either $1$ ($a_i = 0$) or $x^{2^i}$ ($a_i = 1$).  Therefore,
the circuit is constructed by the pairwise controlled constant
multipliers\footnote{Of course, we must classically compute the
  numbers $x^{2^i} (\mathrm{mod} N)$}.

Let $N$ be an $n$ bit number, and $a$ a $2n$ bit number (that
is, $l$ is equal to $2n$ in the above equation.) in the Shor's
factoring algorithm because $a$ is as large as $N^2$. 
$n + 1$ qubits are required as the work-space for the controlled
multiplier and $n + 4$ for the controlled adders. The total number of 
required qubits becomes $5n + 6$.

The circuit is constructed with the $O(l)$ (that is, $O(n)$) pairwise
controlled constant multipliers. The controlled constant multiplier
consists of $O(n)$ controlled constant modular adders. The controlled
constant modular adder consists of 5 controlled constant adders. The
controlled constant adder consists of $O(n)$ XOR (C-NOT) gates. Thus,
the modular exponentiation circuit requires $O(n^3)$ gate.
Detailed are described in Ref \cite{miquel96factoring}. 
It is simulated in $O(n^3 2^ {n-P})$ when there are no overheads for
parallel execution ($2^P$ is the number of processors available in the
simulation system).

\section{Error Model}

\subsection{Decoherence}
We consider the quantum depolarizing channel as the decoherence error
model. In this channel, with probability $1-p$, each qubit is left
alone. In addition, there are equal probabilities $p/3$ that
$\mit\sigma_x\mbox{, }\mit\sigma_y\mbox{, or }\mit\sigma_z$ affects
the qubit.

\subsection{Operational Error}
In general, all of single qubit gates are generated from \emph{rotations}
\[\textnormal{\mathversion{bold}$U_R$}(\theta)= \left(
  \begin{array}{cc}
    \cos\theta & -\sin\theta\\
    \sin\theta & \cos\theta
  \end{array}
\right),\]
and \emph{phase shifts,}
\[\textnormal{\mathversion{bold}$U_{P1}$}(\phi)= \left(
  \begin{array}{cc}
    1 & 0\\
    0 & e^{i\phi}
   \end{array}
  \right)
  \mbox{ and }
   \textnormal{\mathversion{bold}$U_{P2}$}(\phi)= \left(
     \begin{array}{cc}
       e^{i\phi} & 0\\
       0 & 1
     \end{array}
     \right).
     \]
For example, we consider $H_n$ as 
$\textnormal{\mathversion{bold}$U_R$}(\frac{\pi}{4})\textnormal{\mathversion{bold}$U_{P1}$}(\pi)$,
and NOT gate as
$\textnormal{\mathversion{bold}$U_R$}(\frac{\pi}{2})\textnormal{\mathversion{bold}$U_{P1}$}(\pi)$.
The simulator represents inaccuracies by adding small deviations to
the angles of rotation $\theta$ and $\phi$. Each error angle is drawn
from Gaussian distribution with the standard deviation ($\mit\sigma$).

\section{Preliminary Experiment}
We describe the simulation environment and some experiments about
basic quantum circuits.
\subsection{Simulation Environment}
We have developed the simulator on the parallel computer, Sun
Enterprise 4500 (E4500). The E4500 has 8 UltraSPARC-II
processors (400MHz) with 1MB E-cache and 10GB memory. The system clock
is 100MHz. The OS is Solaris 2.8 (64bit OS). The simulator is written
in a C language and the compiler that we use is Forte Compiler 6.0. The
compiler option ``\texttt{-xO5 -fast -xtarget=ultra2 -xarch=v9}''. We
use the solaris thread library for multi-processor execution. Under
this environment, if we use an in-place algorithm, \emph{30-qubit
  quantum register states can be simulated.}

\subsection{Quantum Fourier Transform}

\begin{table}[htbp]
  \begin{center}
    \leavevmode
    \caption{QFT execution time (sec).}
    \label{tab:qft}
    \begin{tabular}{|c|l||r|r|r|r|}
      \hline
      \lw{Qubits} & \lw{Algorithm} & \multicolumn{4}{|c|}{Num. of
        Procs} \\ \cline{3-6}
      & & 1 & 2 & 4 & 8 \\ \hline
      \lw{20} & Circuit & 26.08 & 7.25 & 5.01 & 5.33 \\ \cline{2-6}
      &         FFT     &  1.21 & 0.92 & 0.72 & 0.53 \\ \hline
      \lw{22} & Circuit & 124.78 & 66.96 & 38.03 & 23.40 \\ \cline{2-6}
      &         FFT     &  5.01 & 3.71 & 2.79 & 1.83 \\ \hline
      \lw{24} & Circuit & 643.02 & 331.98 & 183.01 &137.7 \\ \cline{2-6}
      &         FFT     &  20.00 & 12.61 & 8.40 & 5.84 \\ \hline
      \lw{26} & Circuit & 2745.56 & 1469.73 & 799.57 & 526.82 \\ \cline{2-6}
      &         FFT     & 113.29 & 73.08 & 48.39 & 32.84 \\ \hline
      \lw{28} & Circuit & 12597.8 & 6738.13 & 3661.51 & 2338.19 \\
      \cline{2-6}
      &         FFT     & 567.19 & 319.16 & 205.98 & 142.01 \\ \hline
      \lw{29} & Circuit & 31089.6 & 16790.6 & 9189.68 & 5811.49 \\ \cline{2-6}
      &         FFT     &  1232.16 & 697.68 & 423.00 & 286.29 \\ \hline
    \end{tabular}
  \end{center}
\end{table}

Table \ref{tab:qft} shows the QFT execution time by the simulator
using the QFT-circuit and (classical) FFT algorithm.  
The numerical error value is ranged from $10^{-15}$ to $10^{-14}$.
Recall that $2^P$ be the number of processors available in the simulation
system. The FFT algorithm requires $O(n 2^{n - P})$ steps and the QFT
circuit requires $O(n^2 2^{n - P})$ steps for the $n$-qubit quantum
register, if there are no overheads for parallel execution.
The execution time is increased in exponential order in proportional
to $n$. The execution time of the FFT is about 20 $\sim$ 30 times as
fast as that of the circuit.  
Both the execution time are decreased when the number of processors
are increased. The speedup-ratios on 8-processor execution are about $4
\sim 5$.  The reason why the speedup-ratios on 8-processor execution
are not $8$ is that the parallel execution has some overheads that
single processor execution does not have. The parallel execution
overheads are operating system overheads (multi-threads creation,
synchronization, and so on), load imbalance, memory-bus saturation,
memory-bank conflict, false sharing and so on. For small-size
problems, the ratio of overheads to the computation for parallel
execution is relatively large and speedup-ratios on multi-processor
execution may be less than $4$. The decoherence and operational errors
experiment for the QFT is described in Section \ref{sec:exp}.

\subsection{Hadamard Transform}
\begin{table}[htbp]
  \begin{center}
    \leavevmode
    \caption{HT execution time (sec).}
    \label{tab:htbp}
    \begin{tabular}{|c|r|r|r|r|}
      \hline
      \lw{Qubits} & \multicolumn{4}{|c|}{Num. of
      Procs} \\ \cline{2-5}
      & 1 & 2 & 4 & 8 \\ \hline
      {20} & 2.38 & 1.18 & 0.76 & 0.40 \\ \hline
      {22} & 10.85 & 5.73 & 3.20 & 1.35 \\ \hline
      {24} & 46.94 & 24.96 & 13.40 & 9.58 \\ \hline
      {26} & 205.81 & 109.97 & 58.83 & 38.71 \\ \hline
      {28} & 887.40 & 467.71 & 253.82 & 167.31 \\ \hline
      {29} & 2027.9 & 1081.1 & 592.08 & 395.81 \\ \hline
    \end{tabular}
  \end{center}
\end{table}

Table \ref{tab:htbp} shows the Hadamard Transform (HT) execution time
by using the circuit. The HT circuit requires $O(n 2^{n-P})$ steps for
the $n$-qubit quantum register.  The speedup-ratio on 8-processor
execution becomes about 5.

\subsubsection{Effect of Errors}
We have investigated the decrease of the $|0 \rangle \langle 0|$ term
in the density matrix for the 20-qubit register.

\noindent{\bf Decoherence Errors}\\
\begin{figure}[htbp]
  \begin{center}
    \resizebox{0.7\textwidth}{!}{\rotatebox{-90}{\includegraphics*{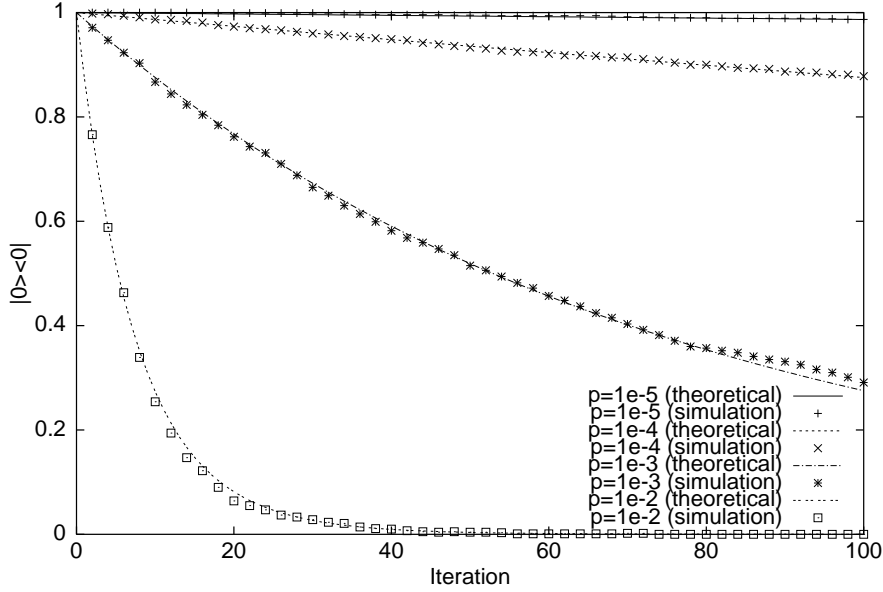}}}
  \end{center}
  \caption{Decrease of the $|0 \rangle \langle 0|$ term in the density
    matrix (20 qubits).}
  \label{fig:20d}
\end{figure}
We have analyzed decoherence in the HT circuit on the depolarizing
channel.
Of course, the simulation deals with pure states. Therefore,
the experiments were repeated 10000 times and we use the average values.
Each experiment uses different initial random seed. The start state of
the quantum register is $|00\ldots0\rangle = |0\rangle$. The HT
circuit is applied to the quantum register over and over. The x-axis
in the Figure \ref{fig:20d} shows the even iteration number. 
If there are no errors (i.e., the error probability is 0) and the
number of iteration is even, the state remains $|0\rangle$ and $|0
\rangle \langle 0|$ term in the density matrix remains 1. 
Figure \ref{fig:20d} shows how decoherence errors degrade the $|0
\rangle \langle 0|$ term. The noise degrades the $|0 \rangle \langle
0|$ term significantly if the error probability is greater than
$10^{-3}$. When the error probability is $10^{-2}$, the $|0\rangle
\langle 0|$ term is decreased in exponential order in proportional to
the number of iterations.

In this easy case, we can compute $|0 \rangle \langle 0|$ term in the
density matrix theoretically. First, consider the 1 qubit case.
Let $p$ be the error probability and $\rho_k$ be the density matrix
after the HT circuit is applied to the quantum register $k$ times. The
density matrix $\rho_{k+1}$ is calculated as follows.
\[
\rho_{k+1} =  (1-p)H\rho_{k}H^{*} +
\frac{p}{3}\sigma_xH\rho_kH^*{\sigma_x}^*\\
 + \frac{p}{3}\sigma_yH\rho_kH^*{\sigma_y}^*
+\frac{p}{3}\sigma_zH\rho_kH^*{\sigma_z}^*\mbox{.}
\]
When the start state of the quantum register is $|0\rangle$ and $k$ is 
even. $\rho_k$ is calculated as follows,
\[
\rho_{k} = {\displaystyle\frac{1}{2}}\left(
  \begin{array}{cc}
    1 + (1-\frac{4}{3}p)^k & 0 \\
    0 & 1 -(1-\frac{4}{3}p)^k \\
  \end{array}
\right)\mbox{.}
\]
In the $n-$qubit case, we can calculate the density matrix
similarly when the start state of the quantum register is
$|0,\ldots,0\rangle$ and $k$ is even. $|0 \rangle \langle 0|$ term of
$\rho_k$ is
\[(\frac{1+(1-\frac{4}{3}p)^k}{2})^n\mbox{.}\]
Figure \ref{fig:20d} also shows this theoretical value of 
$|0 \rangle \langle 0|$ term in the density matrix when $p=10^{-5} \sim
10^{-2}$ and $n=20$. We can see that the simulations and the theoretically
computations yield almost the same result. 

\noindent{\bf Operational Errors}
\begin{figure}[htbp]
  \begin{center}
    \resizebox{0.7\textwidth}{!}{\rotatebox{-90}{\includegraphics*{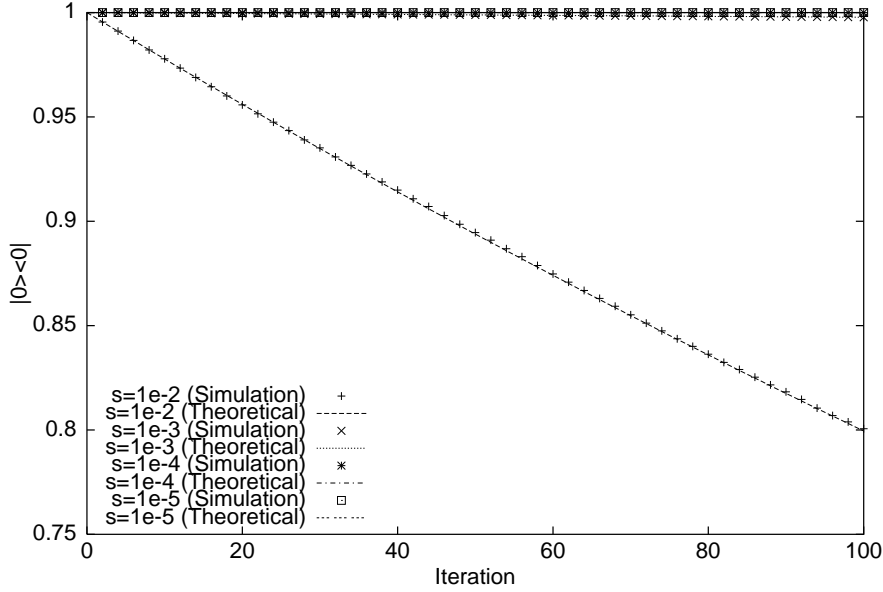}}}
  \end{center}
  \caption{Decrease of the $|0 \rangle \langle 0|$ term in the density
    matrix (20 qubits).}
  \label{fig:20o}
\end{figure}

The simulator represents inaccuracies by adding small deviations to
the two angles of rotations. Since $H =
\textnormal{\mathversion{bold}$U_R$}(\frac{\pi}{4})\textnormal{\mathversion{bold}$U_{P1}$}(\pi)$,
we add small deviations $x$ and $y$ to $\frac{\pi}{4}$ and ${\pi}$
respectively. That is, we use 
$H(x,y)=\textnormal{\mathversion{bold}$U_R$}(\frac{\pi}{4}+x)
\textnormal{\mathversion{bold}$U_{P1}$}(\pi+y)$
as the $H$ gate in this experiment.
$x$ and $y$ are drawn from Gaussian distribution with
the standard deviation ($\mit\sigma$). As mentioned above, the
experiments are executed 10000 times and we use the average
value. Each experiment uses different initial random seed. Figure
\ref{fig:20o} shows how operational errors degrade the $|0 \rangle
\langle 0|$ term when $\sigma = 10^{-5} \sim 10^{-2}$ and $n=20$. The
$|0 \rangle \langle 0|$ term is not affected by the operational error
if $\mit\sigma$ is less than $10^{-2}$.

In this case, we can also compute $|0 \rangle \langle 0|$ term in the
density matrix theoretically. First, consider the 1 qubit case.
Let $\rho_k$ be the density matrix after the HT circuit is applied to
the quantum register $k$ times. The density matrix $\rho_{k+1}$ is
calculated as follows.
\begin{displaymath}
\rho_{k+1} = \int_{-\infty}^{\infty}\!\!\int_{-\infty}^{\infty}\!\!\!H(x,y)
\rho_k {H(x,y)}^* p(x) p(y) dx dy
\end{displaymath}
where $p(z)=\frac{1}{\sqrt{2\pi} \sigma}e^{-\frac{z^2}{2\sigma^2}}$.
When the start state of the quantum register is $|0\ldots0\rangle =
|0\rangle$, $\rho_k$ is calculated as follows,
\[
\rho_{k} = {\displaystyle\frac{1}{2}}\left(
  \begin{array}{cc}
    1 + e^{-\frac{\sigma^2}{4}9k}& 0 \\
    0 & 1 - e^{-\frac{\sigma^2}{4}9k} 
  \end{array}
\right)\mbox{.}
\]
As for the general $n-$qubit case, we can calculate the density matrix
similarly when the start state of the quantum register is
$|0\ldots,0\rangle$ and $k$ is even. $|0 \rangle \langle 0|$ term of
$\rho_k$ is 
\[(\frac{1+e^{-\frac{\sigma^2}{4}9k}}{2})^n \mbox{.}\]

Figure \ref{fig:20d} also shows this theoretical value of 
$|0 \rangle \langle 0|$ term in the density matrix when the standard
deviation $\sigma = 10^{-5} \sim 10^{-2}$ and $n=20$. 
It follows from
the theoretical computation that $|0\rangle \langle 0|$ term is
decreased in exponential order in proportional to the number of
iterations $k$.

\noindent{\bf Both Operational and Decoherence Errors}\\
\begin{table}[htbp]
  \begin{center}
    \leavevmode
    \caption{Combined effects for HT.}
    \label{tab:hod}
    \rotatebox{-90}{
      \scriptsize
      \hspace*{-0.8cm}Decoherence($p$)
    }
    \begin{tabular}{|r||r|r|r|r|}
      \multicolumn{5}{c}{Operational($\mit\sigma$)} \\ \hline
      & 0 & $10^{-5}$ & $10^{-4}$ & $10^{-3}$ \\ \hline \hline
            0 & 1.0000 & 1.0000 & 0.9999 & 0.9977 \\ 
    $10^{-5}$ & 0.9870 & 0.9870 & 0.9849 & 0.9797 \\
    $10^{-4}$ & 0.9010 & 0.9010 & 0.8909 & 0.8780 \\
    $10^{-3}$ & 0.2910 & 0.2790 & 0.2779 & 0.2664 \\ \hline
    \end{tabular}
  \end{center}
\end{table}

Each element of Table \ref{tab:hod} represents the $|0 \rangle \langle
0|$ term of the density matrix after the HT is applied to the
state $|0\rangle$ of a 20-qubit register 10000 times.
The combined effect of two factors may be worse than each factor
alone, that is to say, the effect seems to be the product of each
factor. Table \ref{tab:hod} shows this situation. 

\section{Experiment}
\label{sec:exp}
\subsection{Shor's Factorization Algorithm \cite{shor94algorithms,
    shor97polynomialtime}}
First, we review the algorithm briefly.

\begin{small}
\begin{description}
\item[Input] An $l$ bit odd number $n$ that has at least two distinct
  prime factors.

\item[Output] A nontrivial factor of $n$
  \begin{enumerate}
  \item Choose an arbitrary $x \in \{1, 2, \ldots, n-1\}$
  \item (Classical) Compute $d = \mathrm{gcd}(x, n)$ by using Euclid's
    algorithm. If $d > 1$, output $d$ and stop.
  \item (Quantum) Try to find the order of $x$:
    \begin{enumerate}
    \item Initialize an $l$-qubit register and a $2l$-qubit register
      to state $| 0 \rangle |0 \rangle$.
    \item Apply the HT to the second register.
    \item Perform the modular exponentiation operator. \\
      That is, $|0 \rangle |a \rangle \rightarrow |x^a (\mathrm{mod}
      \ n)\rangle |a \rangle$
    \item Measure the first register and apply the QFT to the second
      register and measure it. Let $y$ be the result.
    \end{enumerate}

  \item (Classical) Find relatively prime integers $k$ and $r$ $(0 < k
    < r < n)$, s.t. $|\frac{y}{2^{2l}} - \frac{k}{r}| \le
    \frac{1}{2^{(2l+1)}}$ by using the continued fraction algorithm.
    If $x^r \not\equiv 1 (\mathrm{mod} n)$ or $r$ is odd or $x^{r/2}
    \equiv \pm 1 (\mathrm{mod} n)$, output "failure" and stop.
    
  \item (Classical) Compute $d_{\pm} = \mathrm{gcd}(n, x^{\frac{r}{2}}\pm1)$
    by using Euclid's algorithm. Output numbers $d_{\pm}$ and stop.
  \end{enumerate}
\end{description}
\end{small}

When the simulator performs all the step-3 operations (not only the
QFT but also the modular exponentiation) on the quantum circuit, 
$5l + 6$ qubits are totally required, as described in the Section
\ref{sec:ari}. Therefore, the simulator can only deal
with 4-bit integer
$n$ ($5l + 6 <= 30 \rightarrow l \le 4$). The 4-bit integer that
satisfies the input property is only 15. We have tried to factor 15 on the
simulator. Beyond our expectation, the modular exponentiation is 
computationally much heavier than the QFT.

\begin{table}[htbp]
  \begin{center}
    \leavevmode
    \caption{Execution time in the Shor's factorization algorithm when
      $n=15$ and $x=11$ (All the quantum operations are executed on
      the circuit).}
    \begin{tabular}{|c|c|}
      \hline
      Modular exponentiation & QFT \\ \hline
      18184 (sec) & 0.64270 (sec) \\ \hline
    \end{tabular}
  \end{center}
\end{table}

The modular exponentiation requires $O(l^3 2 ^{l - P})$ steps
and the QFT on the circuit requires $O(l^2 2 ^{l - P})$ steps
when there are $2^P$ processors available in the simulation system and 
there are no overheads for parallel execution. 
Of course, in the classical computer, modular
exponentiation consists of basic operations such as addition,
multiplication and division. However, these basic operations are not
so heavy in the classical computer, because it has the dedicated
non-reversible circuit (the so-called ALU :arithmetic logic unit). 
This situation suggests that a brand-new fast quantum algorithm for
arithmetic operations are required.
15 is not enough to investigate the behavior of Shor's factoring
algorithm.
To factor much larger number in a reasonable time, the simulator
performs the step-3(c) and the step-3(d) classically. That is, the modular
exponentiation are computed classically and the QFT is computed by the
FFT algorithm in the simulator. In this case, the simulator does not
need to generate the first register. Therefore, the simulator can
factor about $14 \sim 15$-bit integers (for example, 23089).

The factoring algorithm successes with the probability greater than 
\begin{eqnarray*}
\hspace*{-0.4cm}
\mathrm{Prob_{succ}}(n) & = & p_{step2} + (1 - p_{step2}) p_{step3\sim4}\\
       & =&  (1 - \frac{\phi(n)}{n-1}) +
        \frac{\phi(n)}{n-1} \cdot
        (\frac{1}{2}\cdot\frac{4}{\pi^2} \frac{e^{-\gamma}}{\log\log n})
\end{eqnarray*}
where $p_{step2}$ means the probability that the step-2 successes
and $p_{step3\sim4}$ means the probability that step-3 and the step-4
success and $\gamma$ is the Euler constant $\phi(n)$ is the Euler
number of $n$. If the above algorithm is repeated
$O(1/\mathrm{Prob_{succ}}(n))$ times, the success probability can be
as close to 1 as desired.

We choose an $n = p q$ where $p$ and $q$ are prime numbers. This kinds
of integers are chosen in an RSA cryptosystem because it is believed
that it is hard to factor such integers easily. $\phi(n) = (p - 1) (q
- 1)$ for such integers. We have experimented with several RSA-type
$14 \sim 15$-bit integers.

The simulator repeats the above algorithm until a nontrivial factor of
$n$ is found. The simulator records the number of iterations. 
The experiment is executed 100 times and we use the average of
these recorded iterations. We compare the simulation values with
the theoretical number of needed iterations
(i.e.,$1/\mathrm{Prob_{succ}}(n)$).
The results are shown in the Table \ref{tab:hs1}. 
Theoretical values (\textbf{Theoretical}) are about only $2
\sim 4$ times as large as simulation values
(\textbf{Original}). Although much more simulations are required, the
theoretical values seem to be fairly good.

\begin{table}[htbp]
  \caption{Number of needed iterations of Shor's factoring algorithm.}
  \label{tab:hs1}
  \begin{center}
    \leavevmode
    \begin{tabular}{|c||c|c|c|}
      \hline
      \lw{$n$} & \multicolumn{3}{c|}{Num. of Iterations} \\
      \cline{2-4}
      & \lw{Theoretical} & \multicolumn{2}{c|}{Simulation} \\
      \cline{3-4}
      & & Original & Improved \\ \hline
      $21311(=211\cdot101)$ & 15.79 & 6.690 & 1.760 \\
      $21733(=211\cdot103)$ & 15.85 & 8.990 & 2.356 \\
      $22999(=211\cdot109)$ & 16.00 & 6.360 & 1.730\\
      $22523(=223\cdot101)$ & 15.88 & 5.480 & 1.770\\
      $22927(=227\cdot101)$ & 15.91 & 3.790 & 1.470\\
      $22969(=223\cdot103)$ & 15.94 & 8.050 & 2.070\\
      $23129(=229\cdot101)$ & 15.92 & 7.133 & 1.636 \\
      \hline
    \end{tabular}
  \end{center}
\end{table}

As suggested in Ref \cite{shor97polynomialtime}, the algorithm is
optimized so as to perform less quantum computation and more
(classical) post-processing.

\begin{enumerate}
\item \textit{Neighbor $y$ Check}

  If we do not find the relatively prime integers $k$ and $r$ by
  using the continued fraction algorithm, it is wise
  to try $y \pm 1$, $y \pm 2$.

\item \textit{GCD Check}

  Even if $x^r \not\equiv 1 \ (\bmod \ n)$, try to compute
  $d_{\pm} = \mathrm{gcd}(n, x^{\frac{r}{2}}\pm1)$. 

\item \textit{Small Factor Check}

  If $x^r \not\equiv 1 (\bmod \ n)$, it is wise
  to try $2r$, $3r\ldots$. This is because if $\frac{y}{2^{2l}}
  \approx \frac{k}{r}$, where $k$ and $r$ have a common factor, this
  factor is likely to be small. Therefore, the observed value of 
  $\frac{y}{2^{2l}}$ is rounded off to $\frac{k'}{r'}$ in the lowest terms.

\item \textit{LCM Check}

  If two candidates for $r$, that is $r_1$ and $r_2$, have been
  found, it is wise to test $\mathrm{lcm}(r_1, r_2)$ as a
  candidate $r$.
\end{enumerate}

We have tested how much the algorithm is improved by these
modifications. The results are also shown in Table \ref{tab:hs1}
(\textbf{Improved}). The number of iterations are reduced to about
$1/5 \sim 2/5$.
The detailed effect of the improved algorithm is described 
in Table \ref{tab:ieffect}.

\begin{table}[htbp]
  \caption{Detailed effect of improved algorithm}
  \label{tab:ieffect}
  \begin{center}
    \leavevmode
    \begin{tabular}{|c||c|c|c|c|}
      \hline
      \lw{$n$} & \multicolumn{4}{c|}{Ratio of Success/Failure} \\
      \cline{2-5}
      & 1(Neighbor) & 2(GCD) & 3(SF) & 4(LCM) \\ \hline
      $21311$ & 27/9 & 52/19 & 12/4  & 3/4  \\
      $23129$ & 27/9 & 52/19 & 12/4  & 3/4  \\
      $22999$ & 37/6 & 47/79 & 13/8 & 2/58 \\
      $22969$ & 41/8 & 22/82 & 31/22 & 1/28 \\
      $22927$ & 25/3 & 35/49 & 18/2  & 1/28 \\
      $22523$ & 37/6 & 45/76 & 18/22  & 7/54 \\
      \hline
    \end{tabular}
  \end{center}
\end{table}

Each element of Table \ref{tab:ieffect} represents $s/f$ where
$s$ means the number of success iterations and $f$ means the 
number of failure iterations. For example, about $n=23129$, the first
optimization, ``Neighbor Check'' is performed for $27 + 9 = 36$
iterations and the candidate of the order is found successfully in
$27$ iterations. It seems that the second optimization ``GCD Check''
works well for all the $n$ that we have experimented with. From this
result, we can see that even if $x^r \not\equiv 1 (\bmod \ n)$,
$d_{\pm} = \mathrm{gcd}(n, x^{\frac{r}{2}}\pm1)$ often become the
factor of $n$. 
That is, even if the candidate $r$ is not equal to $ord(x)$ (an order
of $x$), there is the possibility that $\mathbf{N} \ni \exists a > 1,\
a \cdot r = ord (x)$. In this case, the following equation holds when
$r$ is even. 
\begin{eqnarray*}
\hspace*{-0.6cm}
  0(\bmod \ n) & \equiv & x^{ord(x)} - 1\\
  & \equiv & (x^{r} - 1) (x^{(a-1)r}+x^{(a-2)r}+\ldots+1) \\
  & \equiv & (x^{r/2} - 1) (x^{r/2} + 1)
  (x^{(a-1)r}+x^{(a-2)r}+\ldots1) 
\end{eqnarray*}
Thus, there is the possibility that $n$ and $x^{\frac{r}{2}}\pm1$
have a common non-trivial factor.

\subsection{Effect of Errors}

We have analyzed decoherence and operational errors in the QFT
circuit.
\begin{figure}[htbp]
  \begin{center}
    {\rotatebox{-90}{\includegraphics*[width=0.535\textwidth]{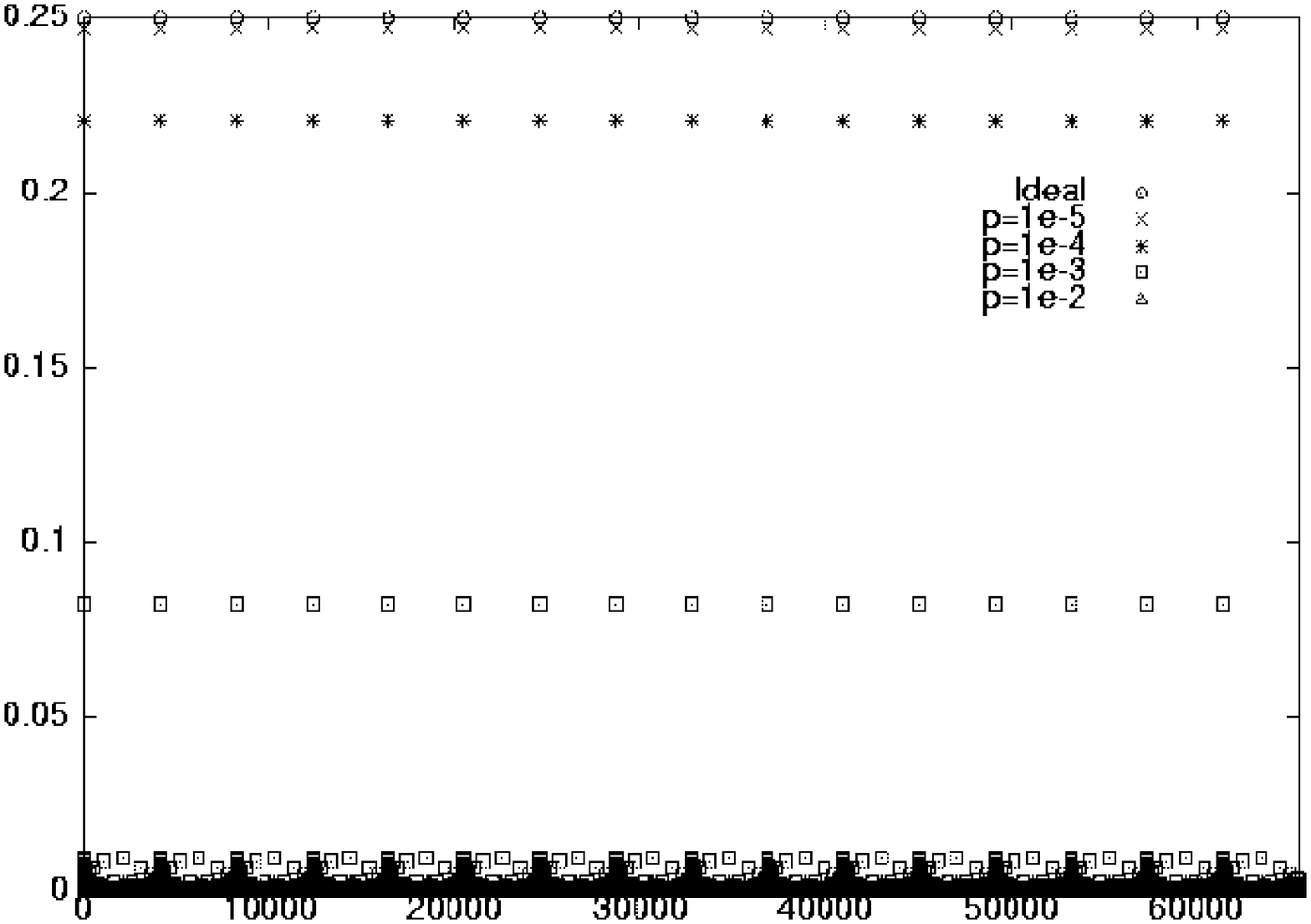}}}
    \begin{scriptsize}
      \begin{tabular}{|c||c|c|c|c|c|}
        \hline
        $p$ & $0$ & $10^{-5}$ & $10^{-4}$ & $10^{-3}$ & $10^{-2}$ \\ \hline
        {Iterations} & 1.9569 & 2.1000 & 2.3201 & 6.0606 & 327.00 \\ \hline
      \end{tabular}
    \end{scriptsize}
    \caption{Amplitude amplification by QFT in the presence of 
      decoherence error (top) and the required number of iterations
      (bottom) (16 qubits).}
  \label{fig:qft16}
  \end{center}
\end{figure}
\begin{figure}[htbp]
  \begin{center}
    {\rotatebox{-90}{\includegraphics*[width=0.5\textwidth]{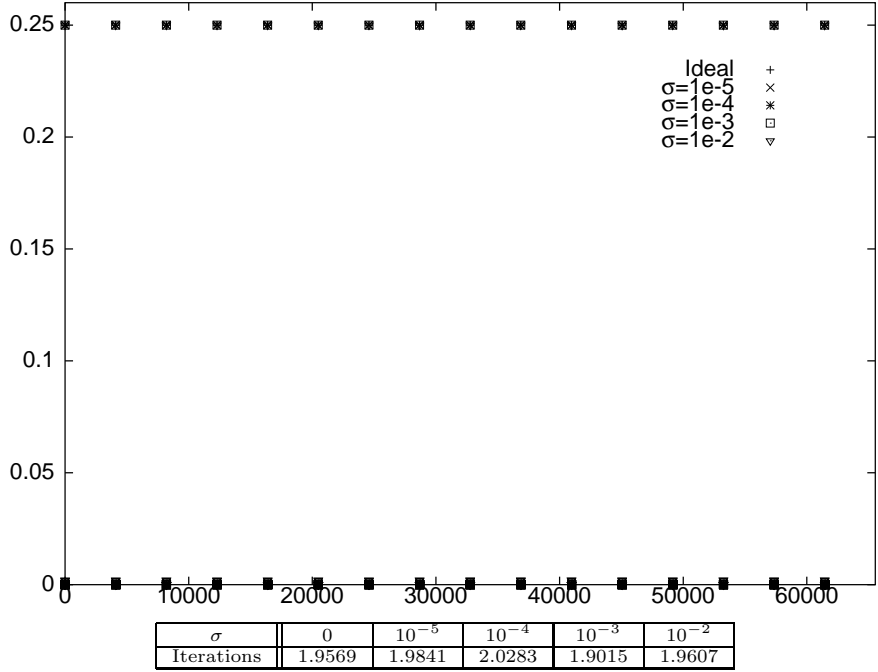}}}
    \begin{scriptsize}
      \begin{tabular}{|c||c|c|c|c|c|}
        \hline
        $\mit\sigma$ & $0$ & $10^{-5}$ & $10^{-4}$ & $10^{-3}$ &
        $10^{-2}$ \\ \hline
        {Iterations} & 1.9569 & 1.9841 & 2.0283 & 1.9015 & 1.9607 \\
        \hline
      \end{tabular}
    \end{scriptsize}
  \caption{Amplitude amplification by QFT in the presence of 
    operational error (top) and the required number of iterations
    (bottom) (16 qubits).}
  \label{fig:qft16o}
  \end{center}
\end{figure}
\noindent{\bf Decoherence Errors}\\
We assume that each qubit is left intact with probability $1-p$
and it is affected by each of the error operators
$\mit\sigma_x,\mit\sigma_y,\mit\sigma_z$ with the same probability
$\frac{p}{3}$ each time the register is applied by the controlled
rotation gate $\mbox{\boldmath $R_d$}$.
Figure \ref{fig:qft16} shows the amplitude amplification phase by the
QFT circuit on the depolarizing channel in Shor's factorization
algorithm (Step 3 (d)) when $n=187$ and $x=23$. The y axe in the Figure
\ref{fig:qft16} shows the amplitude. The experiment is executed 1000
times and we use the average. If the error probability is greater than
$10^{-3}$, it is hard to use the QFT circuit for the purpose of period
estimation.

\noindent{\bf Operational Errors}

The simulator represents inaccuracies by adding small deviations to
the angles of rotations of $\mbox{\boldmath $R_d$}$. We consider $H_n =
\textnormal{\mathversion{bold}$U_R$}(\frac{\pi}{4})\textnormal{\mathversion{bold}$U_{P1}$}(\pi)$,
and NOT gate $ =
\textnormal{\mathversion{bold}$U_R$}(\frac{\pi}{2})\textnormal{\mathversion{bold}$U_{P1}$}(\pi)$.
The simulator also represents inaccuracies by adding small deviations to
these angles of rotations.  The error is drawn from Gaussian
distribution with the standard deviation ($\mit\sigma$). As mentioned
above, the experiment is executed 1000 times and we use the
average. Figure \ref{fig:qft16o}
shows the amplitude amplification phase by the QFT in the Shor's
factorization algorithm (Step 3(d)) when $n=187$ and $x=23$.  
It seems that the period extraction by using the QFT is not affected
by the operational error.

\noindent{\bf Both Operational and Decoherence Errors}\\
We investigate the combined effect of operational and decoherence
errors. Table \ref{tab:qft16od} shows the result. Each element of table
represents the \emph{fidelity}. The fidelity is defined as the inner
product of the correct state and the simulated state with errors. 

The combined effect of two factors may be worse than each factor
alone, that is to say, the effect seems to be the product of each
factor. However, when the decoherence rate is relatively higher, the
small-deviation operational error can improve the results contrary to
our expectations. When the size of register is large, the decoherence
probability even greater than $10^{-3}$ drops the fidelity
significantly.

\begin{table}[htbp]
  \vspace*{-0.5cm}
  \begin{center}
    \leavevmode
    \caption{Combined effects for QFT (16bit)}
    \label{tab:qft16od}
    \rotatebox{-90}{
      \scriptsize
      \hspace*{-0.8cm}Decoherence($p$)
    }
    \begin{tabular}{|r||r|r|r|r|r|}
      \multicolumn{6}{c}{Operational($\mit\sigma$)} \\ \hline
      & 0 & $10^{-5}$ & $10^{-4}$ & $10^{-3}$ & $10^{-2}$ \\ \hline \hline
        0    & 1.0000  & 0.9999 & 0.9999 & 0.9999 & 0.9998 \\
    $10^{-5}$ & 0.9880 & 0.9840 & 0.9860 & 0.9880 & 0.9848 \\
    $10^{-4}$ & 0.8837 & 0.8897 & 0.8827 & 0.8801 & 0.8980 \\
    $10^{-3}$ & 0.3287 & 0.3399 & 0.3332 & 0.3209 & 0.3363 \\ 
    $10^{-2}$ & 0.0027 & 0.0015 & 0.0019 & 0.0017 & 0.0031 \\ \hline
    \end{tabular}
  \end{center}
  \vspace*{-0.4cm}
\end{table}

\subsection{Grover's Search Algorithm \cite{grover96fast}}
\begin{figure}[htbp]
  \begin{center}
    \includegraphics*[width=0.7\textwidth]{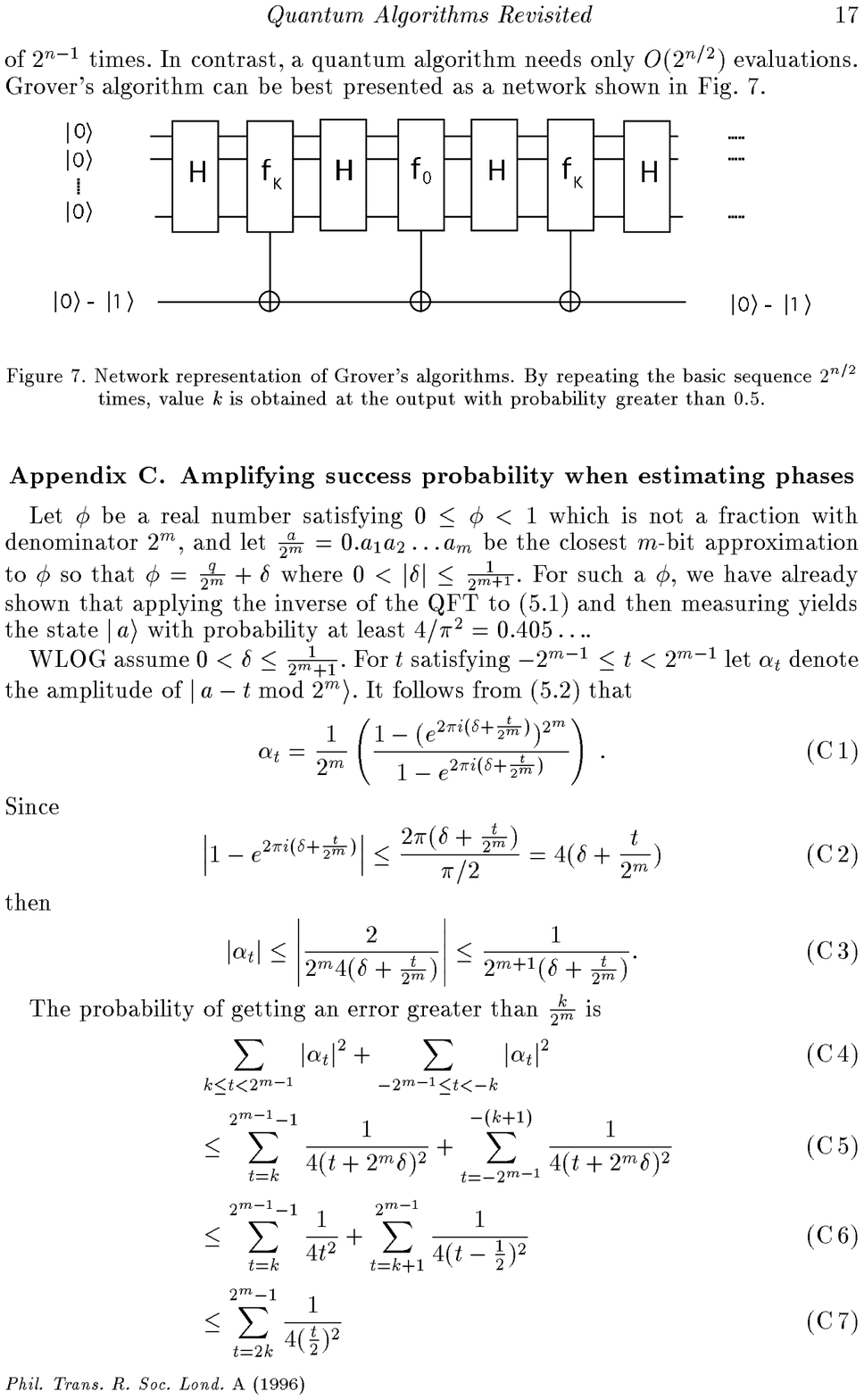}     
    \caption{The circuit of Grover's algorithms.}
    \label{fig:cga}
  \end{center}
\end{figure}

Suppose that a function $f_k : \{0, 1\}^n \rightarrow \{0,1\}$ is an
oracle function such that $f_k (x) = \delta_{xk}$. 
The G-iteration is denoted as $-H_n V_{f_0} H_n V_{f_k}$. 
The sign-changing operator $V_f$ is implemented by using the
$f$-controlled $NOT$ gate and one ancillary bit. 
Figure \ref{fig:cga} shows the circuit of Grover's algorithm. 

\subsubsection{Effect of Errors}

\begin{figure}[hbtp]
  \vspace*{-0.5cm}
  \begin{center}
    \resizebox{0.7\textwidth}{!}{\rotatebox{-90}{\includegraphics*{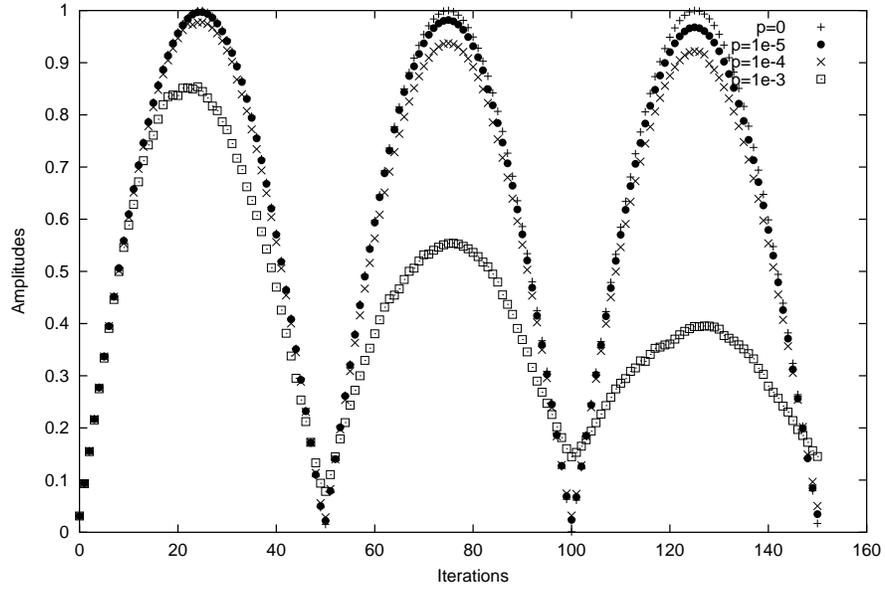}}}
  \end{center}

  \caption{Decrease of the amplitude of the correct element 
    in the presence of decoherence errors (10 qubit).}
  \label{fig:g10d}

\end{figure}
\begin{figure}[hbtp]

  \begin{center}
    \resizebox{0.7\textwidth}{!}{\rotatebox{-90}{\includegraphics*{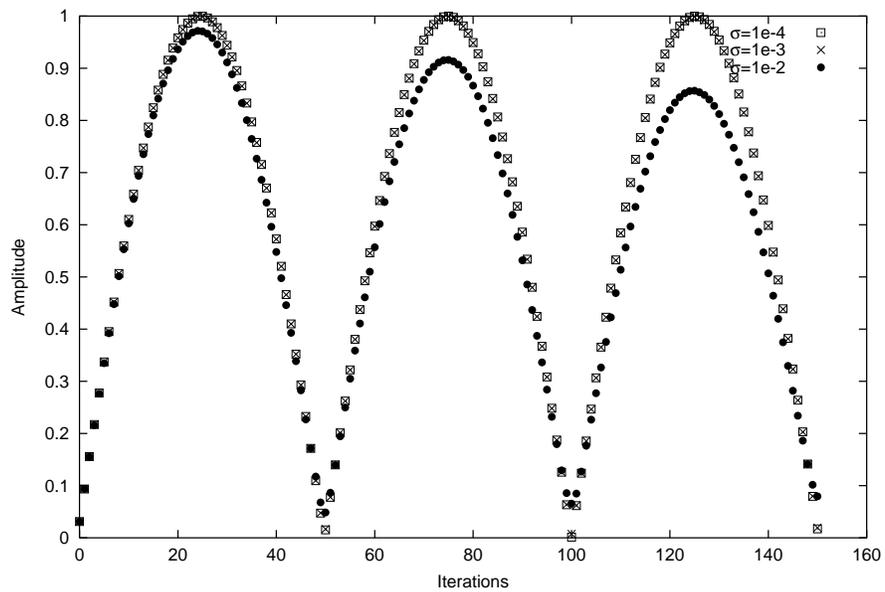}}}
  \end{center}
 \vspace*{-0.1cm}
  \caption{Decrease of the amplitude of the correct element 
    in the presence of operational errors (10 qubit).}
  \label{fig:g10o}

\end{figure}

We have analyzed the impacts of decoherence and operational errors in
the circuit of Grover's algorithm.
We assume depolarizing channel
that each qubit is left intact with probability $1-p$ and it
is affected by each of the error operators $\mit\sigma_x,\mit\sigma_y,
\mit\sigma_z$ with the same probability $\frac{p}{3}$ per G-iteration.
We consider $H_n =
\textnormal{\mathversion{bold}$U_R$}(\frac{\pi}{4})\textnormal{\mathversion{bold}$U_{P1}$}(\pi)$
and NOT-gate $=
\textnormal{\mathversion{bold}$U_R$}(\frac{\pi}{2})\textnormal{\mathversion{bold}$U_{P1}$}(\pi)$.
The simulator represents inaccuracies by adding small deviations to
the angles of these rotations. Each error angle is drawn from Gaussian
distribution with the standard deviation ($\mit\sigma$).

Figure \ref{fig:g10d} and \ref{fig:g10o} show the impacts of errors
for a 10-qubit register. The experiments were repeated 1000 times and
we use the average values. If there are no errors, plotting the
amplitude of the correct element (that is, $k$) makes a sine
curve. However, the amplitudes are decreased as G-iterations are
repeated in the presence of errors. Figure \ref{fig:g10d} shows the
impacts of decoherence error. We can see that the decoherence error
affects the period of the sine-curve. Figure \ref{fig:g10o} shows the
impacts of operational errors. It seems that the operational error
does not affect the period of the sine-curve.

\section{Related Works}

There are many quantum simulators for quantum circuit model of
computation \cite{obenland98parallel, QDD, omer00, OpenQ}.
QDD\cite{QDD} aims to use Binary Decision Diagram in order to
represent the states of quantum register. QCL\cite{omer00} and
OpenQubit\cite{OpenQ} both use complex number representation of the
quantum states like our simulator. In addition, QCL tries to
establish a high-level, architecture-independent programming
language. The Obenland's simulator \cite{obenland98parallel} is based
on an actual physical experimental realization and it uses parallel
processing like our simulator. 
Although it runs on the
distributed-memory multi-computers, our simulator runs on the
shared-memory multi-computers.  Therefore, in our simulator, there is
no need to distribute and collect the states of the quantum
register. In addition, our simulator uses more efficient evolution
algorithms and adopts (classical) FFT algorithms for the fast
simulation of the large-size problems. Our simulator does not depend
on any actual physical experimental realizations because it is not
easy to say which realizations are best at this moment. In other
words, our simulator is more general-purpose.

\section{Conclusion}
We have developed a parallel simulator for quantum computing on the
parallel computer (Sun, Enterprise4500). Up-to \emph{30 qubits} can it
deal with. We have performed Shor's factorization and Grover's
database search by using the simulator, and we analyzed robustness of
the corresponding quantum circuits in the presence of decoherence and
operational errors. If the decoherence rate is greater than $10^{-3}$,
it seems to be hard to use the both quantum algorithms in practice. 
For future work, we will investigate the correlation between
decoherence and operational errors, that is, why small-deviation
operational errors can improve the results when the decoherence rate
is relatively higher. Furthermore, we will try quantum
error-correcting code to fight decoherence and operational errors.
\bibliographystyle{plain}
\bibliography{Paper}
\end{document}